\documentclass[preprint,12pt]{elsarticle}

\usepackage{amssymb}
\usepackage{amsthm}
\usepackage{comment}
\usepackage{url}
\usepackage{booktabs}
\usepackage{makecell}

\usepackage[figuresright]{rotating}
\usepackage{multirow}

\begin{document}

\Large
\textbf{\begin{center}{PRESOL: a web-based computational setting \\ for feature-based flare forecasting} \end{center}}

\normalsize

Chiara Curletto$^{1}$, Paolo Massa$^{2}$, Valeria Tagliafico$^{3}$, Cristina Campi$^{1}$, Federico Benvenuto$^{1}$, Michele Piana$^{1,4}$ and Andrea Tacchino$^{1,3}$ \\

\hspace{-0.5cm}$^1$ MIDA, Dipartimento di Matematica, Universit\`a di Genova, via Dodecaneso 35 16146 Genova, Italy  \\
$^2$ Institute for Data Science, University of Applied Sciences and Arts Northwestern Switzerland, Bahnhofstrasse 6, Windisch, 5210, Switzerland \\
$^{3}$ HoB srl \\
$^4$ Istituto Nazionale di Astrofisica, Osservatorio Astrofisico di Torino, via Osservatorio 20 10025 Pino Torinese Italy \\


\begin{center}
\textbf{Abstract}
\end{center}
Solar flares are the most explosive phenomena in the solar system and the main trigger of the events' chain that starts from Coronal Mass Ejections and leads to geomagnetic storms with possible impacts on the infrastructures at Earth. Data-driven solar flare forecasting relies on either deep learning approaches, which are operationally promising but with a low explainability degree, or machine learning algorithms, which can provide information on the physical descriptors that mostly impact the prediction. This paper describes a web-based technological platform for the execution of a computational pipeline of feature-based machine learning methods that provide predictions of the flare occurrence, feature ranking information, and assessment of the prediction performances.
$$ $$
\noindent
\textbf{keywords:} High-Performance Computing; space weather; Machine Learning; Data Visualization








\section{Introduction}\label{sec:Intro}

Solar flares, the main trigger of space weather \cite{schwenn2006space}, are sudden releases of electromagnetic energy that can be explained according to magnetic reconnection models \cite{tandberg1988physics,piana2022hard}. Solar flare physics presents at least two open issues that currently bother the scientific community of both theoretical physicists and space weather forecasters. The first one is concerned with the effectiveness with which the active Sun is able to accelerate electrons in flares (see \cite{Volpara_2025} and references therein); the second one addresses the problem of solar flare forecasting and nowcasting by means of either model- or data-driven approaches. In the last two decades, data-driven methods for flare forecasting essentially rely on artificial intelligence algorithms \cite{GEORGOULIS2024, georgoulis2021flare}. More specifically, on the one hand deep learning approaches take as input images or videos of solar magnetograms and provide as output binary predictions of the flare occurrence and of the flare energetic content \cite{guastavino2022implementation, 2019SpWea..17.1404C, Li_2025}. Deep learning algorithms are typically easy to train and promising from an operational viewpoint, but the computational burden of their training process is typically high, while their explainability degree is rather low. On the other hand, machine learning methods require a more cumbersome pre-processing phase, based on feature extraction through pattern recognition techniques, but they are characterized by a higher degree of explainability \cite{Benvenuto2018, campi2019feature, 2015ApJ...798..135B, Barnes_2016}.

The present paper describes and discusses ``PRESOL - AI-driven PREdiction of SOLar flares from remote sensing data'', which is, at the same time, a computational pipeline and a technological framework for forecasting flare occurrence using features extracted from solar magnetograms. Specifically, in the PRESOL pipeline
\begin{itemize}
    \item An extended dataset of features extracted from magnetograms provided by the Helioseismic and Magnetic Imager (HMI) \cite{scherrer2012helioseismic} is generated, annotated and appropriately split for training purposes.
    \item A notable amount (currently six) of machine learning algorithms is implemented.
    \item The impact of the different features on the prediction performances of the algorithm is assessed by means of feature ranking methods.
    \item Prediction results are validated by means of a notable amount of scores.
\end{itemize}
One of the nice aspects of PRESOL is that, thanks to its technological framework, it can be executed by means of a web-based, zero-footprint technological platform that also allows the visualization of results and a friendly interactions with the user.

The structure of the paper is as follows. Section 2 describes the pipeline for feature extraction. Section 3 describes the process for the annotation and splitting of the data set. Section 4 provides a quick overview of the machine learning algorithms used for prediction and Section 5 a description of the technological framework for their execution. Our conclusions are offered in Section 6.

\section{The source data: SHARP features}\label{sec:MM}

In order to generate a dataset of features extracted from solar magnetograms we implemented a pipeline that requests and downloads HMI Active Region Patch (SHARP) \cite{2014SoPh..289.3549B} FITS files covering the time range between January 1, 2012 and March 20, 2025. These data were provided by the Joint Science Oprations Center\footnote{\url{http://jsoc.stanford.edu/}} (JSOC) and assessed by means of the \texttt{drms} library \cite{Glogowski2019}. SHARP features were selected based on their physical significance and statistical relevance for solar flare prediction, as established in \cite{2015ApJ...798..135B}. The resulting dataset currently includes 25 descriptors together with their associated uncertainties. These parameters can be broadly grouped into four main categories: 
\begin{itemize}
   \item Parameters illustrating general information, including the observation times (T\_OBS, T\_REC), the unique identifiers of the active regions (HARPNUM), and the corresponding NOAA active region numbers (NOAA\_ARS).
   \item Global magnetic parameters, which consist of quantities derived from the observed magnetic field, such as the total magnetic flux (USFLUX); the mean values and standard deviations of the transverse, vertical, and horizontal magnetic field components (MEANGBT, MEANGBZ, MEANGBH); and parameters related to magnetic topology and electric current distribution, including mean axial current (MEANJZH), vertical current density (MEANJZD), and mean alpha (MEANALP), an indicator of magnetic twist.
   \item Energetic and instability indicators, which encompass estimates of the free magnetic energy available for flare triggering, such as mean and total potential energy (MEANPOT, TOTPOT), free energy associated with horizontal currents (TOTUSJH, TOTUSJZ), and magnetic shear indices, both expressed as a mean value (MEANSHR) and as a threshold-exceeding measure (SHRGT45).
  \item Geometric and morphological parameters, which refer to the size and spatial extent of the active region, including the area in arcseconds squared and the number of pixels (AREA\_ACR, NPIX), the extreme heliographic coordinates of the region (LAT\_MIN, LON\_MIN, LAT\_MAX, LON\_MAX), and aggregated measures of size and complexity (SIZE, NACR).
\end{itemize}
The first release of this resulting 12-year dataset included several invalid data points, characterized by inconsistencies, errors, or non-computable values. To ensure data integrity and prevent these anomalies from compromising model training, both qualitative and quantitative analyses were conducted, which identified and removed corrupted entries such as non-numeric or non-computable values (NaN, inf, -inf), caused by acquisition errors, instrument malfunctions, or extreme solar conditions not handled by the feature extraction codes. We also excluded SHARP features extracted from active regions close to the solar limb, i.e., we retain only those with longitude in [-68°, 68°]. In fact, regions close to the limb exhibit distorted shapes and introduced errors in the flare-association process.

\section{Dataset preparation}

The first step of our feature-based machine learning pipeline for flare forecasting involves the preparation of the features' dataset with the aim to allow the training and validation of a battery of machine learning algorithms. This requires the annotation of each active region according to the possible occurrence of the event and, in the case of flare forecasting, of the flare class; and the implementation of a splitting procedure for the active region set that ensures a realistic class balance for training, validation, and test.

\subsection{Annotation}

We developed an algorithm to label each data point in the HMI feature archive with information about flaring activity, assigning a value of “1” if the corresponding active region (AR) produced a flare and “0” otherwise. We implemented a Python routine to request and download flare event data, covering the same period as the HMI observations, from the Heliophysics Event Knowledgebase\footnote{\url{https://www.lmsal.com/hek/index.html}} (HEK) using the sunpy module Fido\footnote{\url{https://sunpy.org/}}. Flare start time, end time, and intensity are derived from measurements provided by the Geostationary Operational Environmental Satellites (GOES), operated by the National Oceanic and Atmospheric Administration (NOAA). Accurate labeling requires identifying which AR produced each solar flare of GOES class above C1.0, so that events can be correctly linked to the corresponding SHARP magnetograms. The HEK database provides the NOAA active region number for most flares, and, similarly, most SHARP parameter entries in our database include the NOAA number corresponding to the active regions within each HMI patch. This allows us to link SHARP parameters to flares by matching NOAA numbers in the majority of cases. For flares without an associated NOAA number, which are identified only by their heliographic coordinates, we determined the association by checking whether the flare occurred within the SHARP bounding box. To ensure accuracy, we accounted for solar differential rotation to correct for the time difference between the flare occurrence and the magnetogram acquisition.

\subsection{Splitting}

In supervised learning, the dataset must be divided into a training set and a test set. The training set is used to fit the machine learning (ML) model and optimize its parameters, while the test set is used to evaluate the model's predictive performance. This split should ensure that both sets are homogeneous and well-balanced, meaning that the class distribution within each set closely reflects that of the entire dataset. Following an extensive review of the relevant literature, we adopted the methodology described in \cite{guastavino2022implementation}. Data samples corresponding to ARs that produced a flare within the next 24 hours were categorized as TYPE\_X, TYPE\_M, or TYPE\_C, based on the GOES class of the most intense flare produced (X class, M class, and C class, respectively). Conversely, SHARP features extracted from ARs that did not produce flare in the 24h after sample time were classified as: 
\begin{itemize}
    \item Type NO1: if the AR never generated a flare 
    \item Type NO2: if the AR did not generate a flare in the past but originated a flare in the future 
    \item Type NO3: if the AR originated a flare in the 48h before the sample time. 
    \item Type NO4: if the AR did not generate a flare in the 48h before the sample time but originated a flare before the 48h before the sample time. 
\end{itemize}
To train a predictive model and ensure a realistic class balance, it was necessary to split the dataset preserving the label distribution of the types as described above. The split was performed following the methodology outlined in \cite{guastavino2022implementation}, which is based on the following two items: 
\begin{itemize}
    \item To avoid data leakage, samples from the same AR cannot be present in both training and test sets. Therefore, they must be assigned to only one of the two.
    \item  The class distribution of the original dataset must be preserved in both the training and test sets. 
\end{itemize}
The algorithm implementing this splitting procedure randomly selects the active regions to be included in each set, resulting in a different split each run. Due to the constraints on the distribution of types, it is possible that a small number of data points is discarded during the process.

\section{Machine learning algorithms and their validation}

The current release of PRESOL includes six machine learning algorithms for flare forecasting:
\begin{itemize}
    \item Lasso \cite{Tibshirani1996lasso}. 
    \item Logistic regression (Logit) \cite{Wu2009Logit}. 
    \item Random Forest \cite{Breiman2001RF}. 
      \item Multi-layer Perceptron (MLP) as a regressor \cite{Rumelhart1986LearningRB}. 
    \item Multi-layer Perceptron (MLP) as a classifier \cite{Rumelhart1986LearningRB}. 
    \item Support Vector Machine \cite{Cortes1995SupportVectorN}. 
\end{itemize}
For the Lasso, Logistic Regression, Random Forest  and MLP Regressor algorithms to be used as classifiers, we implemented three threshold selection strategies during the training phase. In the first two, the threshold was determined by maximizing either the True Skill Statistic (TSS) or Heidke Skill Score (HSS). The third, a hybrid approach \cite{Benvenuto2018} applied a clustering algorithm to the regression outputs in order to classify the results into two groups (Flare and No Flare). Further, we implemented several metrics to assess the performance of each flare forecasting model. We considered both general metrics designed for classification problems (e.g., Accuracy) and metrics specifically tailored for unbalanced datasets (e.g., True Skill Statistic). The metrics adopted in this study, and described in \cite{Bloomfield2012}, are the True Skill Statistic (TSS), the Heidke Skill Score (HSS), accuracy (ACC), probability of detection (POD), false alarm rate (FAR), and balanced accuracy (BA). Each of these provides specific insights into the model’s performance, which are essential for the user. For instance, POD evaluates the model’s ability to correctly identify positive events, while FAR measures the frequency of false alarms. The ACC, TSS, and HSS scores assess the overall skill of the model, with a maximum value of 1 indicating a perfect prediction. The balanced accuracy score, defined as the average of sensitivity (true positive rate) and specificity (true negative rate), is particularly useful when evaluating models on imbalanced datasets, where one class has significantly fewer samples than the other.
The implementation of the algorithms and metric computations was carried out using Python’s scikit-learn library \cite{scikit}.

We also implemented a feature-ranking algorithm for the SHARP features, namely Recursive Feature Elimination (RFE) algorithm \cite{guyon}. RFE is a recursive procedure that, at each iteration, assigns an importance score to each predictor, removes the least important feature, and retrains the model. This process continues until all the features are ranked.  
To evaluate the algorithms' performance, we split the main dataset 20 times. For each split and each algorithm, we trained and tested a machine learning model, computed the evaluation metrics, and applied the RFE algorithm. This procedure enabled a statistical analysis of the results.

\section{Technological framework}

Probably the most interesting aspect of PRESOL is the web portal designed to integrate and execute the Python pipeline that trains the machine learning methods, and realizes predictions on features extracted from the magnetograms. It provides users with a complete end-to-end workflow, guiding them through model training, testing and prediction, either on the provided dataset (see Section 2 and Section 3) or on custom datasets uploaded by the users.

\begin{figure}[ht!]
    \centering  \includegraphics[width=1\linewidth]{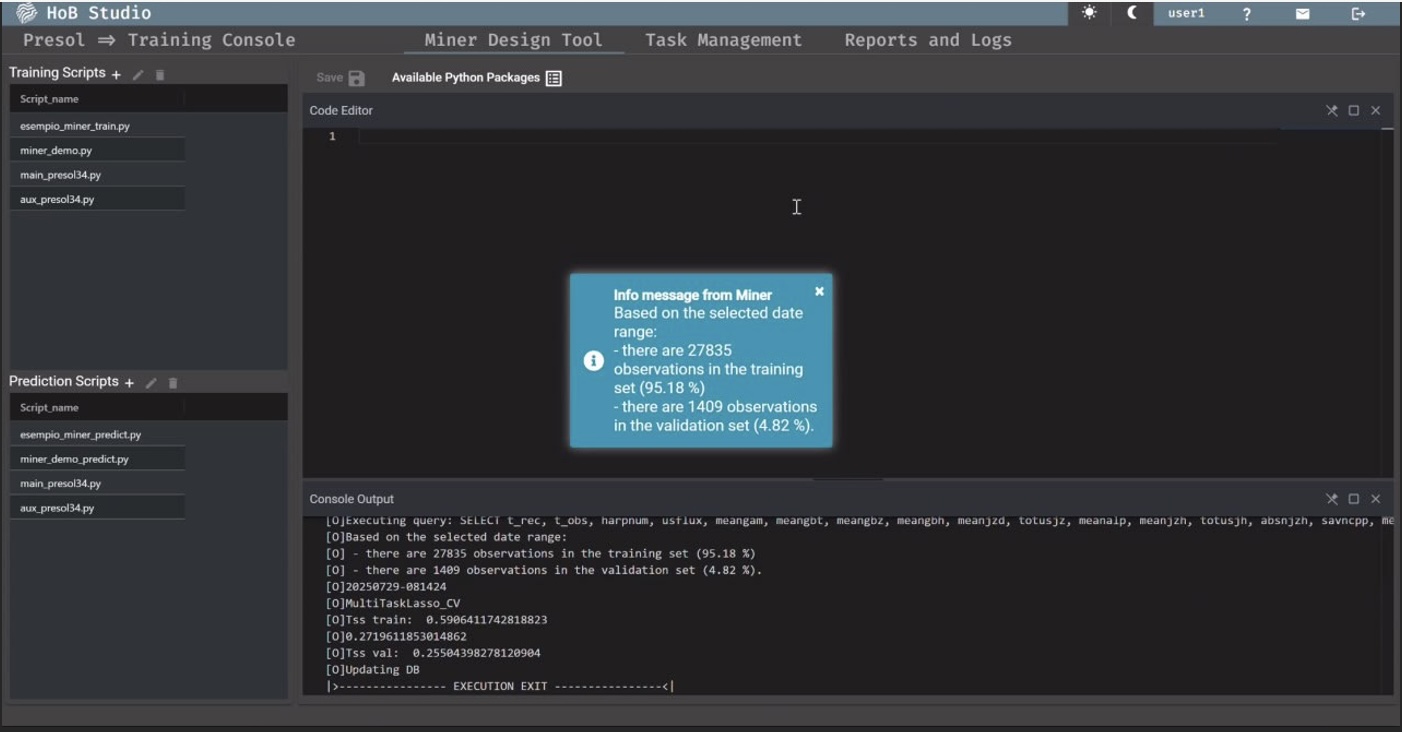}
    \caption{Snapshot of the PRESOL interface with some miners on the left.}
    \label{fig:technology-1}
\end{figure}

\begin{figure}[ht!]
    \centering  \includegraphics[width=1\linewidth]{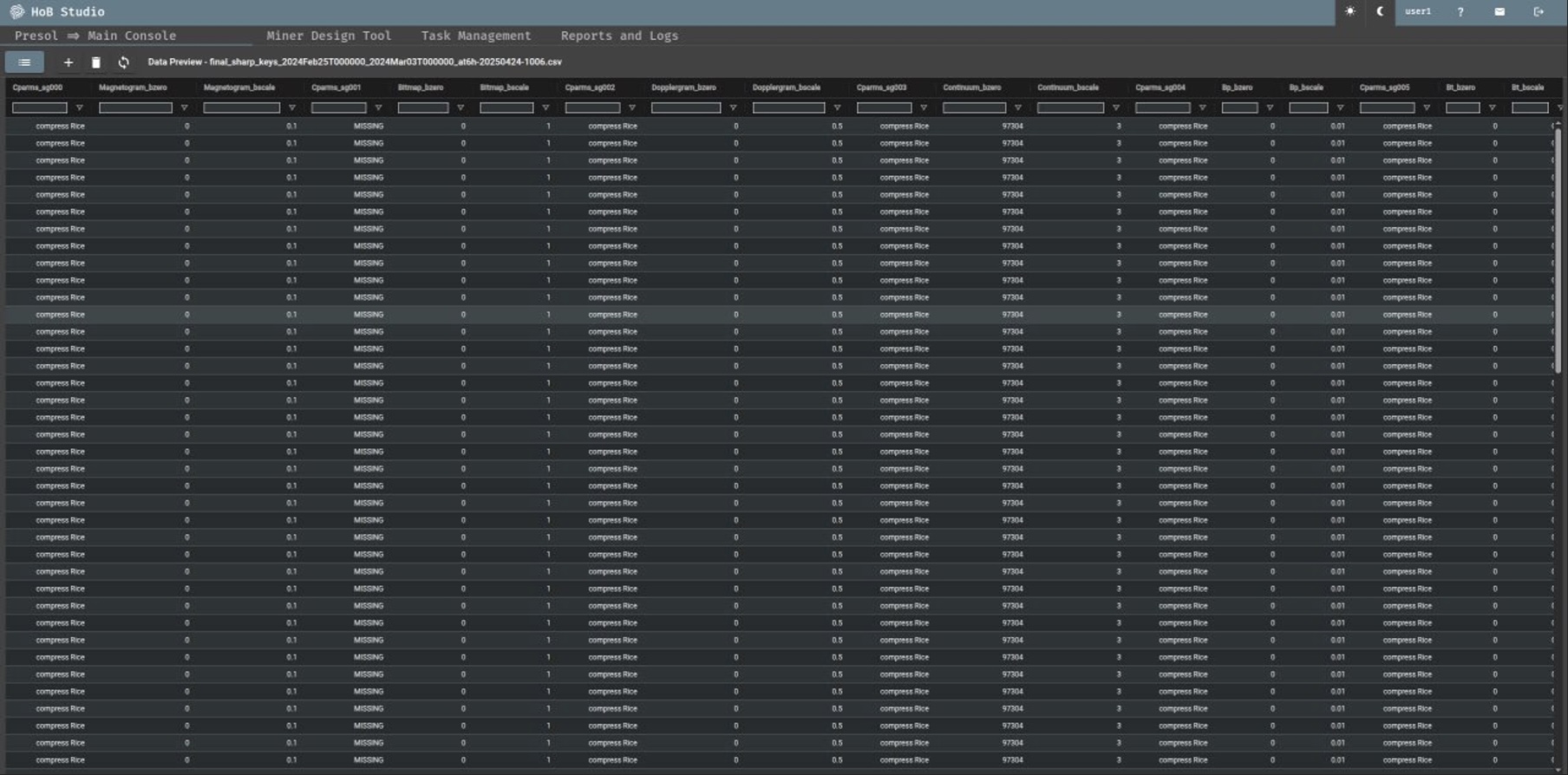}
    \caption{The PRESOL interface previewing all available datasets that can be used for training.}
    \label{fig:technology-2}
\end{figure}

The platform is organized around “miners” (see Figure \ref{fig:technology-1} as an overview of the interface with some miners on the left), defined as sets of techniques and methodologies aimed at extracting useful information from large datasets through automatic or semi-automatic approaches. In this context, miners are essential for analyzing features and identifying patterns, with applications ranging from scientific research to, potentially, operational decision-making.
From the main console, users can: 
\begin{itemize}
    \item Choose between using the dataset provided in this work, or uploading their own labeled dataset in CSV format.
    \item Specify the splitting method for generating the training and test sets.
    \item Select one of the implemented algorithms and customize its parameters through a JSON configuration file.
    \item Launch the training phase.
    \item Apply a previously trained model to new data.
\end{itemize}
A dedicated section of the interface allows users to preview all available datasets before selecting one for training (see Figure \ref{fig:technology-2}). Algorithm configuration is handled via a JSON file that is parsed by the Python backend; the file structure is fixed, and users may modify only parameter values to ensure robustness. Similarly, the choice of the train–test split is defined through a separate JSON file. Three options are available: the methodology of used for the results presented in the present paper \cite{guastavino2022implementation}, chronological splitting, and random stratified splitting, which preserves class distributions in both subsets. Each split type requires specific parameters, which must be provided in the corresponding JSON file.
Once configured, the training process can be initiated. During training, the portal displays progress notifications (e.g., the proportion of data assigned to training and test sets). Upon completion, a compressed archive is generated containing evaluation metrics for both sets, prediction outputs, dataset features, and feature rankings.
In the prediction section, users may apply a previously trained model to either the default dataset or their own uploaded dataset. If the chosen dataset is labeled, the portal automatically evaluates the model and provides the corresponding metrics. A compressed archive with the predictions results can then be eventually downloaded.

\section{Results}\label{sec:Res}
\begin{figure}[ht!]
    \centering
\includegraphics[width=1\linewidth]{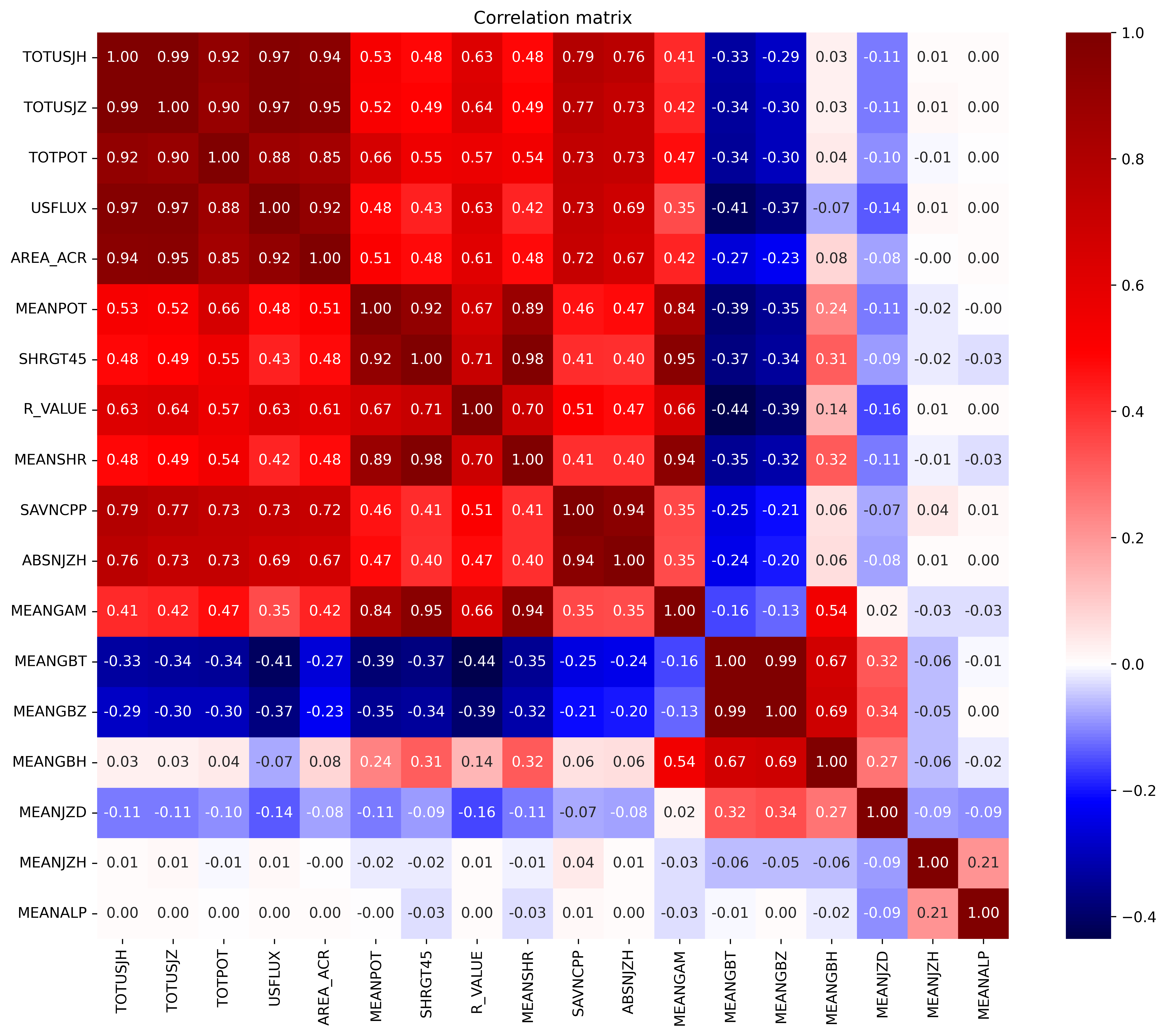}
    \caption{Feature correlation on the whole dataset}
    \label{fig:corr}
\end{figure}

\begin{figure}[ht!]
    \centering  \includegraphics[width=1\linewidth]{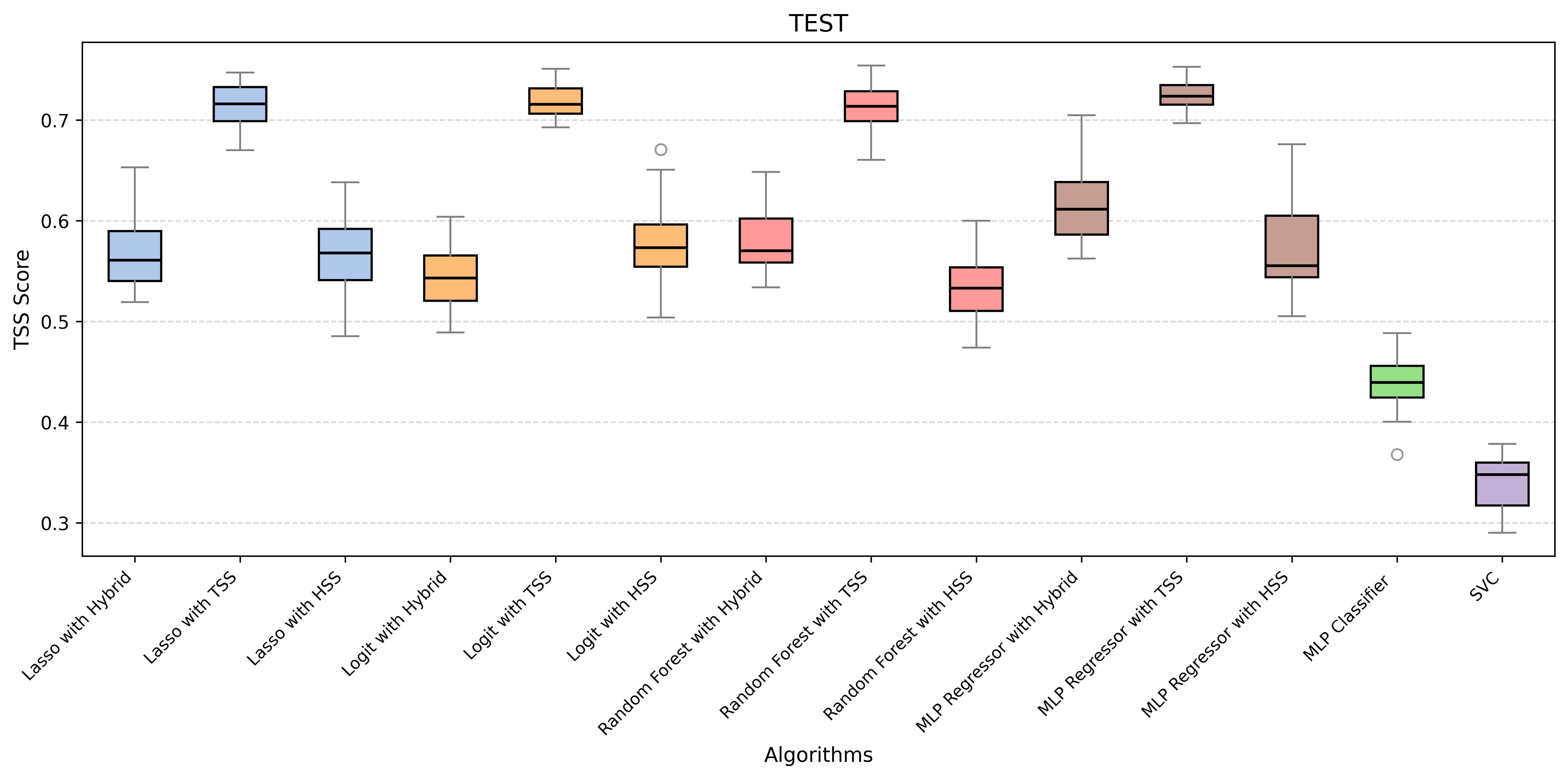}
    \caption{TSS values on the test set using different thresholding optimization.}
    \label{fig:tss_test}
\end{figure}

\begin{figure}[h]
    \centering  \includegraphics[width=1\linewidth]{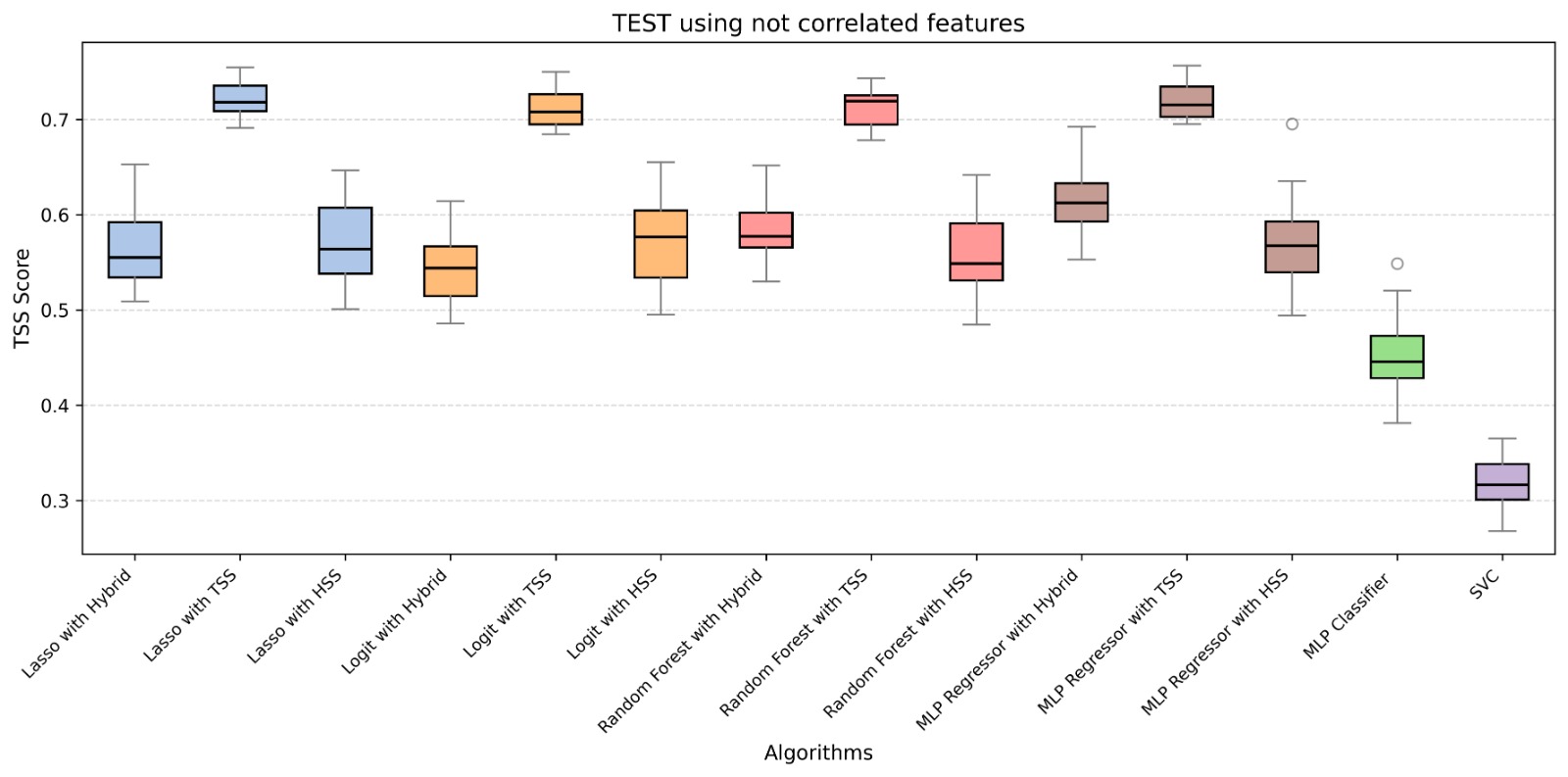}
    \caption{TSS values of the model, without the highly correlated features, on the test set using different thresholding optimization.}
\label{fig:tss_test_reduced}
\end{figure}

Just as an example of application of PRESOL, we considered the dataset currently present in the PRESOL database and applied the battery of machine learning algorithms currently at disposal of the pipeline. Table \ref{tab:metrics_res} shows the results of this application. The table suggests that when the regression algorithms are used as classifiers, the TSS values are rather systematically higher than the HSS ones. Further, using TSS to implement the thresholding step seems the best choice. Anyhow, the standard deviation of the selected thresholds was stable and when thresholding criteria other the TSS are used, the uncertainty on the metrics remains small. The box plots in Figure \ref{fig:tss_test} show these uncertainties in the TSS score for all machine learning methods and all the thresholding criteria used in the experiment. Finally, the entries in Table \ref{tab:ranking_res} are the average feature rankings obtained different rankings across the algorithms: for some features the variability was small (e.g., R VALUES), whereas for others the differences were more pronounced (e.g., USFLUX). This variability is due to a high correlation between the predictors (Figure \ref{fig:corr}). However, several features remained stable across all methods and were consistent with previous findings (e.g., \cite{2015ApJ...798..135B}). We repeated the analysis on a reduced dataset, where high-correlated features (correlation $\geq
90\%$) were removed. The results on the test sets are comparable with the previous ones (see Figure \ref{fig:tss_test_reduced} for the TSS scores and Table \ref{tab:ranking_res_reduced} for the ranking). The feature rankings also became more stable, with the top- and bottom-ranked features showing consistent positions across algorithms. By removing features that were highly correlated with one another, the algorithms were better able to distinguish between the most and least important predictors. For example, both TOTUSJZ and TOTUSJH were initially identified by some algorithms as important, but they were strongly correlated with each other as well as with TOTPOT, USFLUX, and AREA ACR. After removing the correlated features, all algorithms consistently ranked TOTUSJH as the most important predictor, as shown in Tables~\ref{tab:ranking_res} and~\ref{tab:ranking_res_reduced}.

\begin{table}[ht!]
    \centering
    \resizebox{\linewidth}{!}{\begin{tabular}{cccccc}
    \toprule
    {\bf Algorithm} &\makecell{{\bf Threshold} \\ {\bf criterion}} &\makecell{{\bf Threshold value from} \\ {\bf training step}} &{\bf TSS} & {\bf HSS} &\makecell{{\bf Balanced}\\{\bf Accuracy}} \\
    \midrule
    \multirow{3}{*}{Lasso} & Hybrid  &0.28	$\pm$	0.01	&  0.57	$\pm$	0.03	& 	0.55	$\pm$	0.02	& 	0.78	$\pm$	0.02\\
    \cmidrule{2-6}
  
     &  TSS &0.15	$\pm$	0.01	  &0.72	$\pm$	0.02	& 	0.45	$\pm$	0.02	& 	0.86	$\pm$	0.01\\
    \cmidrule{2-6}
      &  HSS &0.28	$\pm$	0.01	& 0.57	$\pm$	0.04	& 	0.55	$\pm$	0.03	& 	0.78	$\pm$	0.02	\\
      \midrule
       \multirow{3}{*}{Logit} & Hybrid & 0.34	$\pm$	0.01	& 0.54	$\pm$	0.03	& 	0.56	$\pm$	0.02	& 	0.77	$\pm$	0.02	\\
        \cmidrule{2-6}
     &  TSS &0.11	$\pm$	0.01	&  0.72	$\pm$	0.02	& 	0.48	$\pm$	0.02	& 	0.86	$\pm$	0.01	\\
      \cmidrule{2-6}
      &  HSS & 0.30	$\pm$	0.02	& 0.58	$\pm$	0.04	& 	0.56	$\pm$	0.02	& 	0.79	$\pm$	0.02	\\
      \midrule
        \multirow{3}{*}{Random Forest} & Hybrid &0.33	$\pm$	0.01 &0.58	$\pm$	0.03	& 	0.56	$\pm$	0.02	& 	0.79	$\pm$	0.02	\\
        \cmidrule{2-6}
     &  TSS &0.14	$\pm$	0.02	& 0.71	$\pm$	0.02	& 	0.51	$\pm$	0.02	& 	0.86	$\pm$	0.01	\\
      \cmidrule{2-6}
      &  HSS & 0.39	$\pm$	0.03	& 0.54	$\pm$	0.04	& 	0.55	$\pm$	0.02	& 	0.77	$\pm$	0.02	\\
      \midrule
       \multirow{3}{*}{MLP Regressor} & Hybrid & 0.32	$\pm$	0.03	& 	0.61	$\pm$	0.04	& 	0.57	$\pm$	0.02	& 	0.81	$\pm$	0.02	\\
           \cmidrule{2-6}
          & TSS &0.13	$\pm$	0.03	& 	0.72	$\pm$	0.02	& 	0.49	$\pm$	0.03	& 	0.86	$\pm$	0.01	\\
           \cmidrule{2-6}
          & HSS & 0.37	$\pm$	0.03	& 	0.57	$\pm$	0.05	& 	0.57	$\pm$	0.02	& 	0.78	$\pm$	0.02	\\   
      \midrule
       {MLP Classifier} & -&- &0.44	$\pm$	0.04	& 	0.51	$\pm$	0.03	& 	0.72	$\pm$	0.02	\\
       \midrule
               SVC & - & -		 & 0.34	$\pm$	0.03	& 	0.44	$\pm$	0.03	& 	0.67	$\pm$	0.01	\\
    \bottomrule
    \end{tabular}}
    \caption{Average $\pm$ sd skill scores on 20 realization of the test set.}
    \label{tab:metrics_res}
\end{table}

\begin{table}[ht!]
    \centering
    \begin{tabular}{ccccc}
    \toprule
\textbf{Algorithms}	&	\textbf{Lasso}	&	\textbf{Logit}	&	\textbf{Random Forest}	&	\textbf{SVC}	\\ \midrule
TOTUSJH	&	1	&	5.05	&	1.6	&	1	\\	\cmidrule{2-5}
R VALUE	&	2.05	&	1	&	2.3	&	14.25	\\	\cmidrule{2-5}
MEANGBZ	&	7	&	3.25	&	9.5	&	4.5	\\	\cmidrule{2-5}
TOTUSJZ	&	13	&	10.35	&	2.1	&	2	\\	\cmidrule{2-5}
AREA ACR	&	3.2	&	8.6	&	4.3	&	12.5	\\	\cmidrule{2-5}
MEANGBH	&	6.65	&	5.65	&	8.7	&	7.65	\\	\cmidrule{2-5}
USFLUX	&	18	&	6.75	&	4.8	&	3	\\	\cmidrule{2-5}
SAVNCPP	&	4.05	&	12.95	&	6.5	&	10.45	\\	\cmidrule{2-5}
MEANSHR	&	7.2	&	6.25	&	15.4	&	7.75	\\	\cmidrule{2-5}
TOTPOT	&	10	&	10.5	&	6.4	&	11.5	\\	\cmidrule{2-5}
MEANGBT	&	16	&	3.7	&	13.75	&	5.2	\\	\cmidrule{2-5}
MEANPOT	&	9.85	&	12.5	&	11.2	&	11.05	\\	\cmidrule{2-5}
SHRGT45	&	8.85	&	13.95	&	14.25	&	8.6	\\	\cmidrule{2-5}
ABSNJZH	&	8.1	&	15.3	&	8.95	&	13.35	\\	\cmidrule{2-5}
MEANGAM	&	17	&	5.45	&	16.5	&	8.25	\\	\cmidrule{2-5}
MEANJZH	&	11.45	&	16.8	&	14	&	16.7	\\	\cmidrule{2-5}
MEANALP	&	12.75	&	17.1	&	14.7	&	16.2	\\	\cmidrule{2-5}
MEANJZD	&	14.85	&	15.85	&	16.05	&	17.05	\\	\bottomrule			
    \end{tabular}
    \caption{Average feature ranking on 20 realization of the training set.}
    \label{tab:ranking_res}
\end{table}

\begin{table}[ht!]
    \centering

    \begin{tabular}{ccccc}
    \toprule
{\bf Algorithms}	&	{\bf Lasso}	&	{\bf Logit}	&	{\bf Random Forest}	&	{\bf SVC} \\ \midrule
TOTUSJH	&	1	&	2	&	1.15	&	1	\\ \cmidrule{2-5}
R VALUE	&	2	&	1	&	1.85	&	6.4	\\ \cmidrule{2-5}
ABSNJZH	&	3	&	6.15	&	3	&	2	\\ \cmidrule{2-5}
MEANGBH	&	4	&	3	&	5.8	&	3	\\ \cmidrule{2-5}
MEANGAM	&	8.8	&	4.3	&	4.1	&	4.05 \\	\cmidrule{2-5}
MEANGBT	&	7.9	&	4.7	&	5.1	&	4.95 \\	\cmidrule{2-5}
MEANJZH	&	5.1	&	8.15	&	8.35	&	7.85 \\	\cmidrule{2-5}
MEANALP	&	6.1	&	7.85	&	8.05	&	7.65 \\	\cmidrule{2-5}
MEANJZD	&	7.1	&	7.85	&	7.6	&	8.1 \\	\bottomrule		
    \end{tabular}

    \caption{Average feature ranking, without the highly correlated features, on 20 realization of the training set.}
    \label{tab:ranking_res_reduced}
\end{table}

\section{Conclusions}\label{sec:Conc}

This paper illustrates PRESOL, a web-based, zero-footprint technological platform for the execution of a computational pipeline of feature-based machine learning algorithms for flare forecasting. The platform allows the user to choose training the algorithms by means of either a pre-uploaded data set or a set generated outside the system. In addition, different types of splitting can be selected to generate training and test sets, and several skill scores can be applied to assess pipeline prediction performances. 

It is now needed to validate the effectiveness and efficiency of the PRESOL platform against numerous experiments of flare forecasting. Further, there are probably two possible directions along with PRESOL might be developed. First, the fact that the pipeline has been conceived according to an intrinsically modular design should encourage the inclusion of new algorithms for data splitting, prediction, and feature ranking. Second, PRESOL could represent the first release of a computational platform for flare forecasting with operational potentialities, and its development as a possible warning machine should be considered.

\section*{Acknowledgments}
This paper is supported by the  Fondazione  ICSC, Spoke 3 Astrophysics and Cosmos Observations. National Recovery and Resilience Plan (Piano Nazionale di Ripresa e Resilienza, PNRR) Project ID CN\_00000013 "Italian Research Center on  High-Performance Computing, Big Data and Quantum Computing"  funded by MUR Missione 4 Componente 2 Investimento 1.4: Potenziamento strutture di ricerca e creazione di ``campioni nazionali di R\&S (M4C2-19)'' - Next Generation EU (NGEU). 
PM is supported by the Swiss National Science Foundation in the framework of the project Robust Density Models for High Energy Physics and Solar Physics (rodem.ch), CRSII5\_193716.
CCu, PM, CCa, FB, and MP kindly acknowledge the support of the National Group of Scientific Computing (GNCS-INDAM).



\section*{Declaration of generative AI and AI-assisted technologies in the writing process}
During the preparation of this work the authors used ChatGPT (OpenAI) in order to improve language clarity and grammar. After using this tool, the authors reviewed and edited the content as needed and take full responsibility for the content of the published article.

\bibliographystyle{elsarticle-num}
\bibliography{references}







\end{document}